\documentclass{svjour3}                     

\smartqed  
\usepackage{graphicx}

\begin{document}

\def\be{\begin{equation}}
\def\ee{\end{equation}}
\def\bea{\begin{eqnarray}}
\def\eea{\end{eqnarray}}
\def\orc{\Omega_{r_c}}
\def\om{\Omega_{m}}
\def\E{{\rm e}}
\def\bearst{\begin{eqnarray*}}
\def\eearst{\end{eqnarray*}}
\def\peleven{\parbox{11cm}}
\def\peffec{\peight{\bearst\eearst}\hfill\peleven}
\def\pspace{\peight{\bearst\eearst}\hfill}
\def\ptwelve{\parbox{12cm}}
\def\peight{\parbox{8mm}}

\markboth{Haghi and Rahvar} {Observational Constraint on MOG}


\title{Observational Constraints on the Modified Gravity Model (MOG) Proposed by Moffat: Using the Magellanic System }

\author{ Hosein Haghi and Sohrab Rahvar}

\institute{Hosein Haghi \at Institute for Advanced Studies in Basic
Sciences (IASBS), P. O. Box 45195-1159,
    Zanjan, Iran
\and
 Sohrab Rahvar \at
 Department of Physics, Sharif University of
Technology, P.O.Box 11365--9161, Tehran, Iran }

\maketitle

\begin{abstract}
A simple model for the dynamics of the Magellanic Stream (MS), in
the framework of modified gravity models is investigated. We assume
that the galaxy is made up of baryonic matter out of context of dark
matter scenario. The model we used here is named Modified Gravity
(MOG) proposed by Moffat (2005). In order to examine the
compatibility of the overall properties of the MS under the MOG
theory, the observational radial velocity profile of the MS is
compared with the numerical results using the $\chi^2$ fit method.
In order to obtain the best model parameters, a maximum likelihood
analysis is performed. We also compare the results of this model
with the Cold Dark Matter (CDM) halo model and the other alternative
gravity model that proposed by Bekenstein (2004), so called TeVeS.
We show that by selecting the appropriate values for the free
parameters, the MOG theory seems to be plausible to explain the dynamics of the MS as well as the CDM and the TeVeS models.
\end{abstract}


\section{Introduction}

The asymptotic flattening of rotation curves of disk galaxies has
been explained by invoking still undetected forms of non-baryonic
dark matter \cite{bos81,rub85}. Dark matter in the form of halo, has
been successfully proposed to explain the dynamics of clusters of
galaxies, gravitational lensing and the standard model of cosmology
within the framework of general relativity. Although the currently
favored Cold Dark Matter (CDM) model has proven to be remarkably
successful on large scales \cite{spe06}, dark matter has not yet
been detected after several experimental efforts. Furthermore, high
resolution N-body simulations are still in contradiction with the
observations on subgalactic scales: the simulations predict more
satellites than what is seen \cite{moo99,kly99}, and the implied
spatial distribution of sub-halos is in contradiction with
observation\cite{gre04}. However, newly discovered galaxies in SDSS
and new simulations considering the environmental effect on the
sub-halo abundance have significantly removed this discrepancy
\cite{ishi08a,ishi08b,simon07}.

In order to explain the missing matter of the Universe, other
alternative theories of gravity have been proposed. In these models,
modification of the laws of gravitation can explain the observed
asymptotically flat rotation curve of galaxies without invoking dark
matter. One of the most famous alternative theories is the Modified
Newtonian Dynamics (MOND) which has been introduced by Milgrom
\cite{mil83}. According to this phenomenological theory, the flat
rotation curves of spiral galaxies are explained by modification of
Newton's second law for acceleration below the characteristic scale
$a_{0}=1.2\times10^{-10} ms^{-2}$ \cite{bek84,beg91,san02}.
Recently, TeVeS\footnote{Tensor-Vector-Scalar}, a Lorentz-covariant
version of MOND, has been presented \cite{bek04}. It cannot explain
the internal dynamics and merging of galaxy clusters such as the
``Bullet Cluster'' without invoking special forms of dark matter
\cite{Clo06,ang06}.

Modified Gravity (MOG) is another alternative theory which has been
proposed by Moffat (2005)\cite{Mof94,Mof05,Mof06a}. It is a fully
relativistic theory of gravitation which is derived from a
relativistic action principle involving scalar, tensor and vector
fields. This theory explained successfully a large number of
galactic rotation curves, the mass and thermal profiles of clusters
of galaxies, the recent data of merging clusters, and the CMB
acoustical power spectrum data \cite{Mof06b,Brow06a,Brow06b,Brow07}.

The dynamics of the satellite galaxies and globular clusters around
the Milky Way could be another approach to examine alternative
gravity models \cite{hagh06,hagh09}. In order to discriminate, at
least in principle, between CDM, MOG and MOND theories, recently the
orbital history of Magellanic Clouds (MCs) and 3D Sun's motion in
the Milky Way have been studied by Iorio \cite{ior09,ior09a}. The
consistency of the dynamics of Magellanic Stream ( MS, a narrow band
of neutral hydrogen clouds started from the MCs and oriented towards
the south galactic pole )\cite{wan72,mat74} in MOND theory has been
investigated in our previous work \cite{hagh06}. Here we generalize
that work to test MOG and compare it with the results obtained with
the TeVeS and CDM model.

In this paper, we use a tidal model for the formation and evolution
of the MS. We study the dynamics of MS for different galactic models
and compare the radial velocity with the observational data.
Modified gravitational effect of galactic disk as the luminous part
of the MW is used to study the dynamics of the MS in the context of
the MOG and the TeVeS. The results are compared with those obtained
by including dark matter halo under the conventional Newtonian
gravity. In this work we use six different sets of initial
conditions that have been reported in literature (Table
\ref{tmodel}). For each set of initial conditions we run a large
number of simulations with different galactic potentials and compare
the outcomes to the observation. The paper is organized as follows.
In section \ref{S2} we give a brief review on the dark halo and disk
model of the MW. In section \ref{S3} we introduce the TeVeS and the
MOG models. The modeling and dynamics of the MS are reviewed in
section \ref{S4}. Results and discussion are given in section
\ref{S5}. The paper is concluded in section \ref{S6}.

\section{Galactic model}\label{S2}

We use the Kuzmin potential for galactic disk. The Newtonian
potential for infinitesimally thin Kuzmin disc is given by
\begin{equation}
\Phi_N(R,z) = \frac{-GM}{\sqrt{R^2+(a+|z|)^2}}, \label{eqn:kuzmin1}
\end{equation}
where $a=4.5$ kpc is the scale length and $M=1.2 \times 10^{11}
M_{\odot}$ is the mass of the disk \cite{read05}.

In order to compare the results with the prediction of CDM theory,
we add an axisymmetric logarithmic halo whose the gravitational
potential is \cite{eva94}
\begin{equation}
\label{phiL}
 \Phi_G^{(L)}=  -{\frac{1}{2}}{{V_0}^2\log{({R_c}^2 +
R^2 +z^2q^{-2})}},
\end{equation}
where, $R_c$ is the core radius, $V_0$ is the asymptotic velocity
and $q$ is the flattening parameter: $q=1$ represents a spherical
halo and $q \neq1$ gives an elliptical halo. Different total masses
of the Milky Way in the form of halo give rise to different values
of $V_0$. Therefore, there are different values of $V_0$ in the
literature: $V_0=161$ kms$^{-1}$ that is used by Jonston et al
(1999) and Law et al (2005) \cite{john99,law05}, $V_0 =175$
kms$^{-1}$ that Read and Moore (2005)\cite{read05} used to reproduce
the tidal feature of Sagittarius, or $V_0=210$ kms$^{-1}$ that
Kochanek (1996)\cite{koch96} obtained from an analysis of the
orbital motion of Galactic satellites. In agreement with Helmi
(2004)\cite{helm04} and Ruzicka et al. (2007)\cite{ruz07}, we set
$V_0 = 185$ kms$^{-1}$ and $R_c=12$ kpc which well reproduced the
kinematics of the Magellanic Stream.

We choose the logarithmic potential for several reasons. First, the
relatively small number of input parameters of Eq.\ref{phiL} makes
the numerical calculations faster. Secondly, the logarithmic
potential was employed in a recent study of dwarf galaxy streams to
investigate the MW dark matter halo \cite{helm04} , and thus the
application of the same formula allows for comparison of our
results. In addition, it allows for the investigation of non
spherical model of the Galactic halo. Finally, the flattened
logarithmic halo model is very nearly identical to the potential of
Kuzmin disk in the deep-MOND regime (see section \ref{S3}).

\section{ MOG and TeVeS}\label{S3}

MOG consists of three theories of gravity called the nonsymmetric
gravity theory (NGT), the metric-skew-tensor gravity (MSTG) theory,
and the scalar-tensor-vector gravity (STVG). MOG has been proposed
by Moffat to explain the rotation curves of galaxies, clusters of
galaxies and cosmology without dark matter
\cite{Mof94,Mof05,Mof06a,moffat96,moffat07a,moffat09b,moffat08,moffat09c,moffat07b}.
It was shown that a parameter-free version of STVG can be obtained
from an action principle \cite{moffat09a}. Good fits to
astrophysical and cosmological data have been obtained with this
more recent version of STVG. An important feature of the NGT, MSTG,
and STVG theories is that the modified acceleration law for weak
gravitational fields has a Yukawa-shape force added to the Newtonian
acceleration law. However, MOG has encountered problems in the solar
system \cite{ior08}. In the weak field approximation limit, STVG,
NGT and MSTG produce similar results. From the field equations
derived from the MOG action, one can obtain the modified Newtonian
acceleration law for weak gravitational fields as
\cite{Mof06a,Brow07}

\begin{equation}
g_{M}=\frac{G(r)M(r)}{r^2}\label{MOGacc}
\end{equation}

\begin{equation}
G(r)= G_N
\times\left\{1+\alpha(r)\left[1-e^{-r/r_0}\left(1+\frac{r}{r_0}\right)\right]\right\},\label{GN}
\end{equation}
where $G(r)$ is the effective gravitational coupling constant, $G_N$
is the Newtonian gravitational constant, $M$ is the baryonic mass,
and $\alpha(r)=\sqrt{M_0M^{-1}}$. Because of the large
galactocentric distance of LMC, we use the point mass approximation
for the total stellar mass (disk and bulge). The masses of the disk
and the bulge are $M_{disk}=10^{11}M_{\odot}$, $M_{bulge}
=3.4\times10^{10}M_{\odot}$, yielding a total baryoinc mass of
$M=1.34\times10^{11}M_{\odot}$ \cite{law05}. Moreover, $M_0(r)$ and
$r_0(r)$ are two scaling parameters that vary with distance and
determine the coupling strength of the vector field to baryonic
matter and the range of the force, respectively. These parameters
are determined by the equations of motion for effective scalar
fields derived from an action principle \cite{Mof06a}. In order to
calculate the MOG dynamics, we have to phenomenologically obtain
$M_0$ and $r_0$: this determines $G(r)$. It is postulated that $M_0$
and $r_0$ give the magnitude of the constant acceleration as
\begin{equation}
\label{M0r0}g_0=\frac{GM_0}{r_{0}^2}. \label{postulate}
\end{equation}
We assume that, for galaxies and galaxy clusters this acceleration
is determined by $g_0=cH_0$, where $H_0 = 100h$ $kms^{-1}Mpc^{-1} $
is the current measured Hubble constant and $h = (0.71 \pm 0.07)$.
This gives $g_0 = 6.90 \times 10^{-10} ms^{-2}$.

Analyzing the galaxy rotation curves, a satisfactory fit to LSB and
HSB galaxy data is obtained with the parameters
$M_0=9.6\times10^{11} M_\odot$ and $r_0=13.9$ kpc, whereas in the
dwarf galaxies smaller than 12 kpc, the best fit value of parameters
obtained as $M_0=2.4\times10^{11} M_\odot$ and $r_0=9.7$ kpc, and
for the satellite galaxies, the parameters are $M_0=4.6\times10^{13}
M_\odot$ and $r_0=111.3$ kpc \cite{Brow06a}. An empirical fitting of
$M_0$ versus $r_0$ for the wide range of spherically symmetric
systems, from the solar system scale to clusters of galaxies has
been obtained and depicted in Fig. 2 of \cite{Brow07}. The
modifications to gravity in Eq. (\ref{MOGacc}) would be canceled by
decreasing $M_0$ and increasing $r_0$ (i.g. $M_0\rightarrow 0$ and
$r_0\rightarrow\infty$). These parameters are scale dependent; thus,
they are not to be taken as universal constants. MOG is not arised
from a classical modification, but from the equations of motion of a
relativistic modification to general relativity.

According to TeVeS model, the physical metric near a quasi-static
galaxy is given by the same metric as in general relativity with
just a little change: the Newtonian potential for known matter,
$\phi_N$ replaced by the total $\Phi$ which comes from two parts,
\begin{equation}
 \Phi = \Phi_{N} + \phi_s , \label{pot}
\end{equation}
where, $\phi_s$ is a potential due to a scalar field. The added
scalar field plays the role of the dark matter gravitational
potential and the corresponding modified acceleration is
$g=g_N+g_s$. The scalar field potential for the Kuzmin disk is as
following \cite{bra95}
\begin{equation}
\phi_s \simeq \frac{(MGa_0)^{1/2}}{2}\ln(R^2+(|z|+a)^2).
\label{eqn:mondpotential}
\end{equation}
From Eqs. (\ref{eqn:kuzmin1}), (\ref{pot}) and
(\ref{eqn:mondpotential}), we calculate the accelerations of a test
object orbiting the galaxy.

Because the Poisson's equation is non-linear in MOND, the strong
equivalence principle is violated \cite{bek84}, and consequently the
internal properties and the morphology of a stellar system are
affected both by the internal and external fields. This effect is
specific for MOND; it significantly affects non-isolated systems
and, in principle, it should be taken into account.

In the presence of an external field, the total acceleration, which
is the sum of the internal $a_i$ and external $a_e$ accelerations,
satisfies the modified Poisson equation \cite{bek84}
\begin{equation}
 \nabla.[\mu(\frac{|\textbf{a}_i+\textbf{a}_e|}{a_{0}})(\textbf{a}_i+\textbf{a}_e)] \simeq 4\pi G \rho, \label{mon1}
\end{equation}
where ${\bf a_e}$ approximately is constant,
$\textbf{a}_i=\nabla\phi$ is the non-external part of the potential,
and $\rho$ is the density of the star cluster. For spherical system
one can approximately write equation \ref{mon1} as $
\textbf{a}_i\mu(|\textbf{a}_i+\textbf{a}_e|/a_{0})=\textbf{a}_N$.
The EFE is indeed a phenomenological requirement of MOND and it was
postulated by Milgrom (1983) to explain the dynamical properties of
nearby open clusters in MW. Equation \ref{mon1} is only an
approximate way to take into account the external field effect, in
order to avoid from solving the modified Poisson equation with an
external source term $\rho_{ext}$ on the right-hand side. The EFE
allows high velocity stars to escape from the potential of the Milky
Way \cite{fam07,wu07}, and implies that rotation curves of spiral
galaxies should fall where the internal acceleration becomes equal
to the external acceleration \cite{gen07,wu08}.

In the three body problem of LMC-SMC-MW interaction, the external
field of MW plays an important role in the internal interaction of
LMC and SMC as MONDian EFE ( for more details see the paper by Iorio
\cite{ior09}). Since the used model for magellanic system in this
paper is the single cloud model (i.e. we neglect the presence of
SMC), there is no mutual interaction between Clouds, and therefore
we do not consider the EFE in the orbital motion of LMC. In the next
section using the RK4 technique we solve numerically the equations
of motion.

\section{Modeling and Dynamics of Magellanic System }\label{S4}

The Magellanic Stream is due to past LMC-SMC-MW interaction; it
extends along a curved path as a narrow band of neutral hydrogen
clouds, originating from the MCs and oriented towards the south
galactic pole \cite{wan72,mat74}. The radial velocity and the column
density of this structure has been measured by many groups
\cite{mat87,bru05}. The observed HI radial velocity profile is
measured along the Magellanic Stream in the Galactic Standard of
Rest (GSR) frame (see caption of Table 1). Large-area 21-cm
\footnote{The electron and proton in the hydrogen atom can have
their spins parallel or antiparallel. A transition between these two
states is called a "spin-flip" transition and leads to the emission
of a photon whose wavelength is 21 cm. This is in the radio part of
the electromagnetic spectrum.} radio surveys have produced most of
the information now available about the detailed structure of the
MS. Since its discovery as a long stream of HI gas trailing the MCs,
a number of models have attempted to explain the dynamics and origin
of the MS. Observations show that the radial velocity of the MS with
respect to the Galactic center changes from $0$ kms$^{-1}$ at the
beginning to $-200$ kms$^{-1}$ at the end of the tail.

There are two main explanation for the origin of the MS: the
tidal-stripping \cite{mur80,lin82,gar94,gar96,wei00,con05} and the
ram-pressure model \cite{moo94,hel94,sof94,meu85,mas05,ruz07}.
\begin{table}
\begin{center}
\begin{tabular}{|c|c|c|}
\hline
IC      &3D $v(x,y,z)(km/s)$    & $r(x,y,z)(kpc)$      \\
\hline GN96   &(-5,-226,194)                     & (-1.0,-40.8,-26.8)   \\
      HR94   &(-10.06,-287.09,229.73)           & (-0.8,-40.8,-27.9)\\
      K2     &(-91,-250,220)                    & (-0.8,-41.5,-26.9)   \\
      K1 Mean&$(-86\pm12,-268\pm11,252\pm16)$   & (-0.8,-41.5,-26.9)   \\
      M05    &(-4.3,-182.45,169.8)              & (0,-43.9,-25.1)     \\
      vdM02  &$(-56\pm36,-219\pm23,186\pm35)$   & (-0.8,-41.5,-26.9)   \\
\hline
\end{tabular}                
\end{center}
\caption{The LMC initial conditions (IC) in Galactocentric
coordinates which is a right-handed Cartesian system with the origin
at the galactic center, the galactic plane at $Z=0$ and the sun at
the position $R_{\odot}$= (-7.9,0,0). References: Gardiner \&
Noguchi (1996, GN96), Heller \& Rohlfs (1994, HR94), Kallivayalil et
al. (2006a,b, K1 and K2), Mastropietro et al. (2005, M05), and van
der Marel et al (2002, vdM02). \label{tmodel}}
\end{table}
According to the tidal hypothesis, the interaction of the MCs with
the MW turns out the materials to form the tidal tails emanating
from opposite side of the Clouds. In the tidal scenario, although
the observed radial velocity profile of the stream has been modeled
remarkably well, the smooth HI column density distribution does not
agree with the observations and the expected stars in the stream
have not been observed yet.

The other hypothesis invokes the idea of ram pressure stripping of
gas from the MCs by an extended halo of diffused gas around the
Galaxy. The drag force on the gas between the MCs causes weakly
bound material to escape and form a trailing gaseous stream. This
material after escaping from the MCs falls towards the Galaxy. On
the other hand, also the ram pressure models have their own
problems. The clumpy structure of the Stream can hardly be
reproduced by the process of continuous ram pressure stripping and
can not reproduce the observed slope of the radial velocity profile
along the MS, especially the high negative velocity tip of the
Stream. Naturally, if the gas in the galactic halo has a clumpy
distribution, which seems likely, this would lead to a clumpy MS in
the ram pressure scenario.

In this section we calculate the dynamics of MCs in modified gravity
theories, using their its present positions and velocities as the
initial conditions for the equations of motion. The overall external
force on the MCs is the sum of the gravitational pull by the
Galactic halo and disk, hydrodynamical drag force from the extended
gaseous halo and dynamical friction force from the Galactic halo. In
MOG and TeVeS models, since we have ignored the Galactic halo, there
is no dynamical friction and the main factor in the gas stripping is
the tidal force exerted by the Galactic disk. It should be noted
that, in the absence of extended dark halo, dynamical friction
produces if an object moves through the visible mass distribution of
the host galaxy. There is no dynamical friction for an object that
moves outside the mass distribution. In the case of MCs, since the
Clouds move very far from the center of the MW (i.e. outside the
visible mass distribution), there is no dynamical friction in the
equations of motion. The general equation of motion of the MCs is
\begin{equation}
\label{lmc}
\frac{d^{2}{\vec{r}}_{c}}{dt^{2}}=\frac{\partial}{\partial r_{c}}[
\phi_{G}(r_{c})] + \frac{{\vec{f}}_{c} + \vec{F}_{drag}}{m_{c}},
\end{equation}
where, $\vec{r}_{c}$ is the distance of the MCs from the Galactic
center, $m_c$ is the mass of the Cloud, $\phi_G$ is the
gravitational potential of the Galaxy, $\vec{f}_{c}$ is the
dynamical friction force, and $\vec{F}_{drag}$ is the hydrodynamical
drag force. We adopt the standard form of dynamical friction as
follows \cite{bin87}
\begin{equation}
\vec{f}_{c}= 0.428 \ln\Lambda \frac{Gm_{c}^2}{r^{2}}
\frac{\vec{v}_{c}}{v_{c}},
\end{equation}
where $ \ln\Lambda \sim 3$ is the coulomb logarithm and $v_c$ is the
relative speed with respect to the gaseous halo. Here we model the
MCs with a dense sphere moving through the gaseous halo; thus the
drag force on this sphere is given by
\begin{equation}
F_{drag}= \frac{1}{2}C_d\rho_g v_c^{2}\pi D^2 ,\label{drag}
\end{equation}
where $C_d$ is the coefficient of drag force, $\rho_g$ is the
density of halo and $D$ is the size of the MCs. The drag force plays
no significant role in the orbital motion of LMC \cite{hagh09a}.
Therefore, we will neglect the drag force for the rest of the paper.

One of the problems with modeling the MCs--MW interaction is an
extended high dimensional parameter space. In order to reduce the
parameter space, one can neglect the influence of the SMC in the
LMC-SMC-MW interaction (single cloud model). Lin \& Lynden-Bell
\cite{lin77} showed that such configuration could explain the
existence of trailing tidal stream. Furthermore, Sofue \cite{sof94}
considered the continuous ram pressure stripping in simulation of
the Magellanic system, ignoring the presence of the SMC. Recently,
the LMC--MW interaction without including the action of SMC has been
studied, and the Stream's properties have been successfully
reproduced \cite{mas05}. The assumption of ignoring the SMC is also
motivated by the fact that the SMC's impact on the orbital history
of the LMC is minimal, because of $M_{LMC} \gg M_{SMC}$
\cite{bes07}. Thus, the global dynamics of the LMC alone is
sufficient for discrimination of gravity model. Since the aim of
this study is examining the overall properties of the MS in the
context of MOG and TeVeS, we use the single cloud model and ignore
the presence of SMC.

In the modified gravity models, since we have ignored the Galactic
halo, there is no significant dynamical friction on the MS, and the
main factor in the stripping of the hydrogen from the MCs is the
tidal force exerted by the Galactic disk.

Another simplifying assumption is the elongation of the MS. Although
the formation of the MS is still a subject of debate, both the ram
pressure and the tidal stripping models predict that the orbits of
the Clouds
should trace the MS at least for less than a Gyr in the past. In
other words, the MS, as seen on the sky, is following the MCs and it
is reasonable to assume that the MS is a narrow and long column of
gas which is moving along the same orbit as the MCs
\cite{gar94,gar96,mas05,bes07}. This assumption is motivated by the
work by Johnston et al. \cite{joh99}, who suggested that the tidal
stream acts as "fossil record" of the recent orbital history of
their progenitors and could thereby provide a probe of the galactic
potential. In this way, the mean velocity of gas in the MS will be
the orbital velocity of the MCs. In order to explain this assumption
one can consider the dynamics of nearby satellites around the MW
under the external tidal field \cite{del04,mur99,dol04,ode03,mon06}.
There is a connection among the tidal tails and satellites's orbits.
After the formation and evolution of tidal tails around the
satellite, the tails are aligned with the orbital path only when the
satellite is near the perigalaction of the orbits
\cite{dol04,ode03}. The degree of elongation of the tails along the
satellite orbital path strongly depends on the ellipticity of orbit.
In the circular orbit case, the tails are clear tracer of the
cluster path, but for most eccentric orbits, the tails are strictly
elongated along the orbital path only when the satellite is near the
perigalaction, whereas at the apogalaction they tend to deviate from
the satellite path \cite{mon06}. Differently stated, in the case of
eccentric orbits, the angle between tail and orbit decreases, when
approaching the pericenter and reaches a minimum at the pericenter,
and then increases moving away from it. Since the MCs are at the
perigalaction, we assume that the stream is approximately elongated
along the orbital path, and thus the mean velocity of gas would be
the orbital velocity of the MCs. It must be noted that if the gas in
the galactic halo itself moves, this effect can put the MS into
almost any direction, making it impossible to use its dynamics to
test CDM versus different gravity theories.

In the next section, we will use tidal scenario for the dynamics of
MS obtaining the trajectories of the MCs in various gravitational
models and comparing the radial velocity profile of MS with the
observed data.

\begin{figure}[t]
\begin{center}
\centerline{\includegraphics[width=110mm,height=130mm]{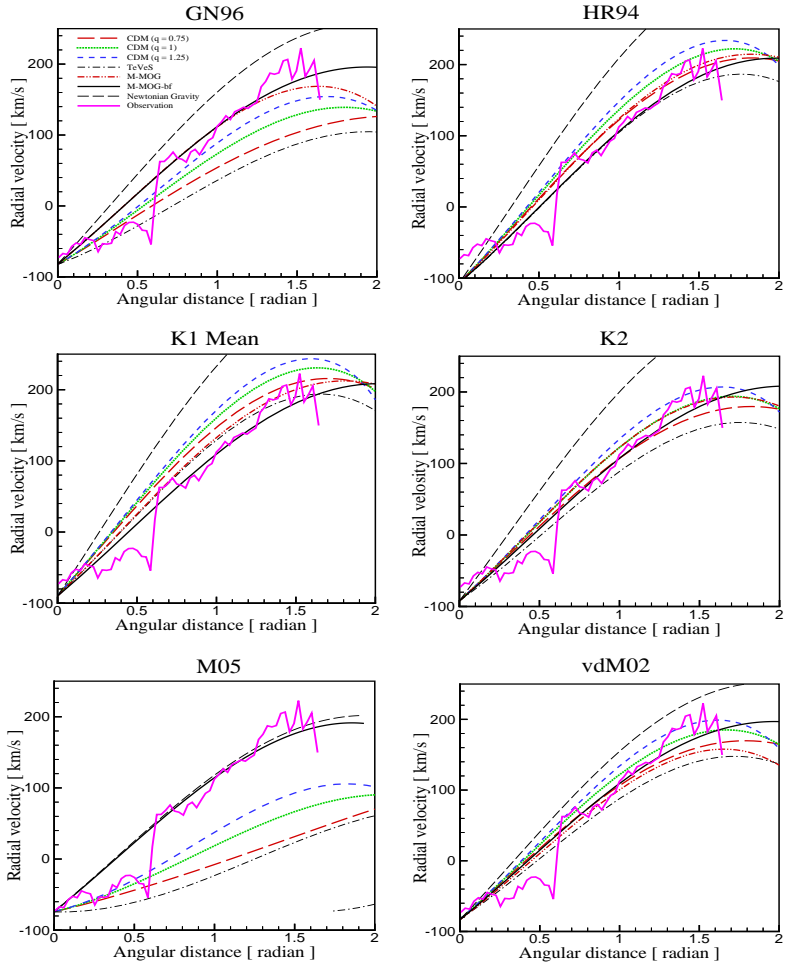}}
\caption{The radial velocity versus the angular distance along the
MS plotted for differen galactic models and compared with the
observational data. The results of the Newtonian gravity model
without dark matter, the CDM halo for different values of $q$, and
the best fit TeVeS model are indicated. The Moffat's modified
gravity predictions in terms of the standard values of $M_0$ and
$r_0$ (M-MOG) and with the best fit values (M-MOG-bf) are also
presented. Each figure corresponds to one of the initial conditions
listed in Table \ref{tmodel}.} \label{V}
\end{center}
\end{figure}

\begin{figure}[t]
\begin{center}
\centerline{\includegraphics[width=80mm,height=110mm]{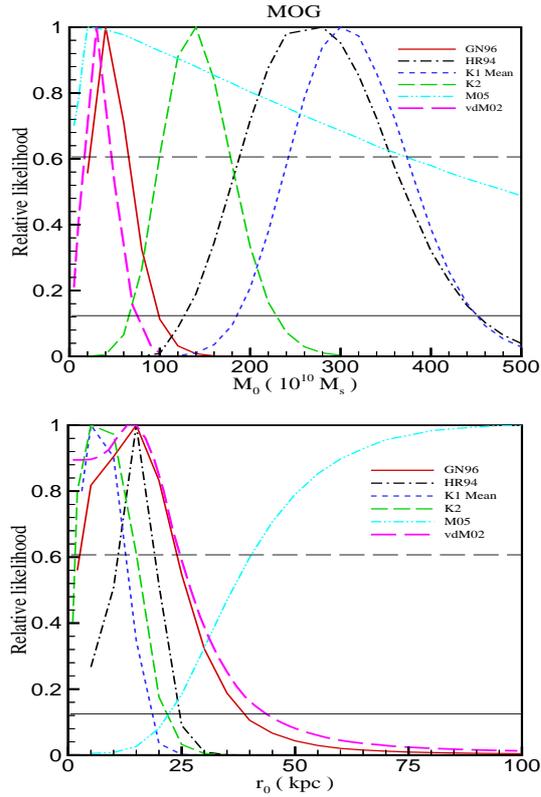}}
\caption{Marginalized likelihood functions of the free parameters of
MOG theory ($r_0,M_0$) for various initial conditions. The location
of picks correspond to the best fit value of parameters. The
intersection of the curves with the horizontal solid and dashed line
give the bound at 1$\sigma$ and 2$\sigma$ confidence levels,
respectively.} \label{likMOG}
\end{center}
\end{figure}

\begin{figure}[t]
\begin{center}
\centerline{\includegraphics[width=90mm,height=65mm]{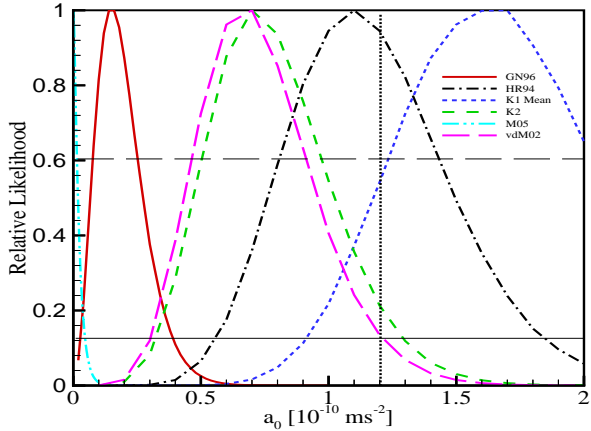}}
\caption{Marginalized likelihood functions of the critical
acceleration in TeVeS, for various set of initial conditions. The
location of the picks correspond to the best fit values of
parameters. The intersection of the curves with the horizontal
dashed and solid line give the bounds at 1$\sigma$ and 2$\sigma$
confidence levels, respectively. The currently accepted value for
$a_0=1.2$ on galactic scales are indicated by vertical dotted line.
The predicted value of $a_0$ in the case of HR94 is in good
agreement with the conventional value $a_0$. Except GN96 and M05,
the other initial condition predicts the acceptable value for $a_0$
at 2$\sigma$ confidence levels.} \label{lik}
\end{center}
\end{figure}

\begin{table*}
\begin{center}
\begin{tabular}{|c|c|c|c|c|c|c|c|}
\hline
        1&work             &$GN96$     &$HR94$   &$K2$    &$K1-Mean$&$M05$    &vdM02       \\
 \hline 2& $\chi^2_{NG}$          &16.4       &63.9     &50.3    &90.7     & 2.5     &13.3        \\
 3&$\chi^2_{MOG}$  &2.06       &1.34     &1.61    &1.71     &2.41     &1.97        \\
 4&$r_0[kpc]$ &$15^{+8}_{-12}$&$15^{+5}_{-5}$&$5^{+12}_{-4}$&$5^{+10}_{-5}$&$110^{+..}_{-60}$& $13^{+11}_{-13}$ \\
 5&$M_0[10^{10}M_{\odot}]$ &$40^{+25}_{-20}$&$280^{+80}_{-100}$&$140^{+40}_{-40}$&$300^{+80}_{-60}$ &$20^{+360}_{-20}$& $30^{+20}_{-15}$ \\
 6&$\chi^2_{TeVeS}$   &2.2        &1.35     &1.8     &2.12     &2.38     &2.39        \\
 7&$a_{0}[10^{-10}ms^{-2}]$ &$0.15^{+0.08}_{-0.05}$& $1.13^{+0.32}_{-0.33}$ & $0.7^{+0.25}_{-0.20}$ & $1.7^{+0.30}_{-0.46}$& $0.01^{+0.01}_{-0.0}$& $0.72^{+0.20}_{-0.25}$ \\
 8&$\chi^2(q=0.75)$&22.9       &2.39     &2.2     &8.11     &87       & 38         \\
 9&$\chi^2(q=1)$   &11.2       &5.58     &2.36    &14.2     &52       &9.8         \\
 10&$\chi^2(q=1.25)$&4.9        &10.1     &4.45    &20.7     &32       & 4.24       \\
  11&$\bar{q}$               &1.7        &0.4      &0.7     &0.2      &2.7      & 1.43       \\
 12&$\chi^2_{CDM_{\bar{q}}}$ &1.61       &1.67     &2.0     &3.1      &2.83     &1.9         \\
\hline
\end{tabular}
\end{center}
\caption{ The best fit parameters of gravitational model and
corresponding minimum values of $\chi^2$ for different initial
conditions (IC). The second row shows the $\chi^2$ of the Newtonian
gravity. The $\chi^2$ and corresponding best fit parameters in MOG
represented in rows of 3-5 as well as TeVeS in rows of 6 and 7. The
error bars are obtained by marginalized likelihood analysis (Figures
2 and 3). The results of logarithmic CDM halo model for different
values of $q$ are shown in rows of 8-10. The last two rows indicate
the best fit value of $q$ and $\chi^2$ for the logarithmic
halo.\label{Result}}
\end{table*}

\section{Results and discussion }\label{S5}

The results of maximum likelihood analysis of the radial velocity
profile of the MS for various gravity models, are presented in this
section. One of the challenging points in calculating the orbital
motion of the MCs is the initial conditions. The trajectory of the
MCs is highly sensitive to the initial conditions, and slight
variation may lead to completely different scenarios \cite{ior09}.
Furthermore, the parameter space of the Magellanic system evolution
is very large \cite{fuj76,mur80,lin82,gar94,gar96}. Due to the
implied numerical burden, doing the fully self-consistent simulation
for each set of initial conditions and different combinations of
parameters is impossible. Recently, an extended analysis of the
parameter space for the interaction of the Magellanic system with
the MW has been done \cite{ruz07}. They have used the genetic search
algorithm and an approximate restricted N-body simulation method.
Also in another orbital analysis, different sets of initial
conditions have been used to determine the dynamics of the LMC
\cite{bes07}.

We numerically integrate the equations of motion in 3D using the
forth-order Runge-Kutta technique with a timestep of 0.5 Myr and
total time of 5 Gyr. In MOG and TeVeS models, the radial velocity
profile of the MS is extracted numerically for various initial
conditions of the MCs listed in Table \ref{tmodel}. In the next step
we use $\chi^2$ fitting to constrain the parameters of the model.
Figure \ref{V} compares the observed data with the theoretical
predictions for various initial conditions. The data points
represent the angular variation of observed radial velocity with
respect to an observer located at the center of Galaxy \cite{bru05}.
The zero angular distance corresponds to the position of MCs center
of mass. The models reproduce radial velocity of the stream as an
almost linear function of angular distance with the high velocity
tip reaching $200$ kms$^{-1}$.

To test the consistency of MOG and TeVeS with observations, we
compare their theoretical predictions to the data directly obtained
from observations and find the best-fitting parameters. Since the
agreement with the data cannot be perfect, we give confidence
intervals for the free parameters of the model using likelihood
analysis. We compute the quality of the fitting through the
least-squares fitting quantity $\chi^2$ defined by
\begin{equation}
 \chi^2\{p_{\alpha}\}=\frac{1}{N}\sum_{i=1}^{N}\frac{(V_{theory}^{i}\{p_{\alpha}\}-V_{obs}^{i})^2}{\sigma_i^2}\label{chi}.
\end{equation}

where $\sigma_i$ is the observational uncertainty in the radial
velocity, $N$ is the number of degrees of freedom
\footnote{$N=n-n_{p_{\alpha}}$, where $n$ is the number of
observational data points and $n_{p_{\alpha}}$ is the number of free
parameters.} and $p_{\alpha}$ is the model parameters. In the case
of MOG, $p_{\alpha}=\{r_0,M_0\}$ and in the case of TeVeS,
$p_{\alpha}=a_0$. Using Eq. (\ref{chi}) we find the best fitted
values of the model parameters that minimize $\chi^2\{p_{\alpha}\}$.

To constrain the parameters of the model, we carry out statistical
analysis using the marginalized likelihood function defined by
\begin{equation}
 \zeta\{p_{\alpha}\}=\xi e^{-\chi^2\{p_{\alpha}\}/2}\label{marg},
\end{equation}
where $\xi$ is a normalization factor. The marginalized likelihood
function of the model parameters are plotted in Figures 2 and 3.
According to Eq. (\ref{marg}), the maximum values of $\zeta$
correspond to the minimum values of $\chi^2$. Therefore, the
location of the peaks correspond to the best fitted values of the
parameters. The main benefit of using the $\zeta$ function is in
finding the $1\sigma$ (68.3 \%) and $2\sigma$ (95.4 \%) confidence
intervals of best-fitting values of the parameters. The best-fitting
values for the parameters of the model at $1\sigma$ and $2\sigma$
confidence intervals for various initial conditions are shown in
Figures 2 and 3. The best-fitting values for the parameters of the
model at the $1\sigma$ confidence interval with the corresponding
$\chi^2$ are presented in Table 2.

The second row of Table \ref{Result} shows the $\chi^2$ of Newtonian
gravity considering the only baryonic matter for Galaxy. As we
expect, the values of $\chi^2$ are very large which means that we
need the dark matter or modification of gravity. The most
interesting result is that the initial condition M05 is almost
compatible with the observation due to the low value of $\chi^2$,
which predicts very small amount of dark matter for the galaxy.

In MOG model we apply the likelihood analysis to find the best
values of the mass scale $M_0$ and range parameter $r_0$. The
minimum values of $\chi^2$ and corresponding best fit parameters are
represented in the rows of 3-5. The error bars obtained by
marginalized likelihood analysis (Fig. \ref{likMOG}). According to
the values of $M_0$ and $r_0$ in section \ref{S3} which were
obtained from fits to satellite galaxies and rotation curves of
galaxies, the reasonable ranges for Magellanic system would be in
the range of $[90 - 5000]\times 10^{10}M_\odot$ for $M_0$ and $[14 -
111]$ kpc for $r_0$. However, the obtained best fit values of $r_0$
are almost in the acceptable range for all sets of initial
conditions, but the $M_0$ values are out of range in the case of
GN96 and vdM02 (see Table \ref{Result}). Furthermore, the values of
$r_0$ and $M_0$ in K1-Mean and M05 do not follow Eq.
(\ref{postulate}) which postulated in MOG theory even considering
the error bars. The case of HR94 is in excellent agreement with
observation, because of the reasonable best fit parameters and the
smaller value of $\chi^2$. Therefore, we can conclude that, the MOG
theory could be compatible with the observational feature of MS by
choosing the appropriate values for free parameters and for special
sets of initial conditions of the MCs.

In the TeVeS model, with the standard value of the acceleration
scale, $a_0=1.2\times 10^{-10}$ ms$^{-2}$, except HR94 which reached
a low $\chi^2$ value, the other initial conditions are not
compatible with observations. However, the value of $a_0$ has been
fixed from rotation curve analysis by Begemann et al.
(1991)\cite{beg91}, it is worth to obtain the value of $a_0$ from
some other independent method. Moreover, among the MOND community
there is no common idea about the value of $a_0$ at different
scales. At different scales like clusters of galaxies or small dwarf
galaxies, it seems the standard value of $a_0$ does not give
acceptable results. At large scales, a lower value of $a_0$, and at
subgaactic scales a larger value of $a_0$ work better
\cite{lok01,nus02,san05,bek08}). In this paper, since we are
studying the dynamics of MCs, i.e. local group scale, it is probable
to find another best-fit value for $a_0$. For this reason, in order
to see the effect of different choice of $a_0$, we allow $a_0$ to
changes and find a best-fit $a_0$ for each initial condition which
gives the better fit with observation.

Figure \ref{lik} depicts the best fit values of $a_0$ for different
initial conditions. Except models GN96 and M05 which prefer a small
value of $a_0$, for the other initial conditions (Table
\ref{Result}), the best fit value of $a_0$ are almost in agreement
with standard value. Thus, in the TeVeS model, GN96 and M05 are
ruled out due to their strange perdition for $a_0$.

In order to compare the results with the CDM models, we apply the
same analysis for the logarithmic halo model. We adopted different
values for halo flattening parameter to deal with prolate, oblate
and spherical halos. According to Table \ref{Result}, the minimum
values of  $\chi^2$ belong to HR94 and K2. They prefer the oblate
and spherical halo models $(q=0.75-1)$. According to the last row,
HR94, K2, and K1-Mean prefer oblate halos whereas, GN96, M05 and
vdM02 prefer the prolate ones. In addition, the minimum values of
$\chi^2$, in the CDM model for GN96 and HR94, are comparable with
MOG and TeVeS, which means that the alternative gravity models can
successfully explained the observational velocity profile as well as
the CDM halo models.

According to the minimum values of $\chi^2$ for different gravity
models in Table 2, for the majority of initial conditions, the MOG
model fit the observations better. However, the difference is not
enough to discriminate between different gravity models.

\section{Conclusion}\label{S6}

In our paper we test alternative gravity models by comparing of the
radial velocity profile of the Magellanic Stream (MS). We
numerically integrated the orbits and dynamics of the Magellanic
Clouds (MCs) within different gravitational models. The MS is
produced by the interaction of the MCs with the Galaxy, and is
considered to be a tracer of the MCs. The drag force plays no
significant role in the orbital motion of the Clouds. In the absence
of the dark halo of the Milky Way, since the Clouds move outside the
visible mass distribution, there is no dynamical friction in the
equations of motion. We used the single cloud model and ignore the
presence of the SMC. The gravitational tidal effect of Galaxy is
assumed as the origin of the MS and it follow the same orbit with
the same dynamics as the LMC. The radial velocity of this structure
is compared with the observation, allowing us to put constraints on
the free parameters of models.

A preliminary inspection of the fits to the Magellanic Stream showed
that for some individual initial conditions, the MOG theory,
choosing the appropriate values of free parameters, $r_0$ and $M_0$,
can be consistent with the observational velocity feature of MS, as
well as TeVeS, and CDM hypothesis. However, due to small difference
in minimum values of $\chi^2$, the discrimination between different
gravity models is impractical.The fits to the Magellanic Stream
based on the parameters $M_0$ and $r_0$ of the older study of STVG,
reveal that the new parameter-free version of STVG \cite{moffat09a}
appears to agree well with the results presented in this paper.

We should point out that while the dynamics of the MS is in a good
agreement with observations, a problem with tidal scenario could be
the lack of corresponding stellar tidal debris in the MS. In TeVeS
and MOG, since the tidal radius is larger than of the Newtonian
case, we expect a Galactic structure with a gas distribution that is
more extended than the stellar component form of the MS. N-body
simulations of tidal stripping of the MS from the MCs in MOG without
dark matter will give a better view of this model and enable us to
compare the density distribution of the gas in the MS with
observations. As an additional result it seems that the initial
condition HR94, is in a good agreement with observation in all
gravitational models.

\begin{acknowledgements}
We thank Alireza Moradi for useful comments on the English of
paper.
\end{acknowledgements}

\end{document}